\documentclass[showpacs,aps,twocolumn]{revtex4}

\usepackage{bm}
\usepackage{amsmath}
\usepackage{graphicx}
\usepackage{subfigure}
\usepackage[usenames,dvipsnames]{color}
\definecolor{darkblue}{RGB}{0,0,196}
\usepackage[colorlinks=true,linkcolor=darkblue,citecolor=darkblue,urlcolor=darkblue]{hyperref}

\usepackage{xcolor}

\usepackage{setspace}
\usepackage{footmisc}
\usepackage[makeroom]{cancel}
\usepackage{comment}
\usepackage{lineno}

\def\be{\begin{equation}}
\def\ee{\end{equation}}
\def\ba{\begin{eqnarray}}
\def\ea{\end{eqnarray}}

\usepackage{graphicx}
\usepackage{amsmath,bbm}
\usepackage{amssymb,bm}
\usepackage{yfonts}

\begin{document}

\title{Predictions for azimuthal anisotropy in Xe+Xe collisions at $\sqrt{s_{NN}}$ = 5.44 TeV using a multiphase transport model}
\author{Sushanta Tripathy}
\author{Sudipan De}
\author{Mohammed Younus}
\author{Raghunath~Sahoo\footnote{Corresponding author: $Raghunath.Sahoo@cern.ch$}}
\affiliation{Discipline of Physics, School of Basic Sciences, Indian Institute of Technology Indore, Simrol, Indore 453552, India}

\begin{abstract}
\noindent
Xe+Xe collision at relativistic energies may provide us with a partonic system whose size is approximately in between those produced by p+p and Pb+Pb collisions. The experimental results on anisotropic flow in Xe+Xe and Pb+Pb collisions should provide us with an opportunity to study system size dependence of $v_2$. In the present work, we have used AMPT transport model to calculate charged particles' $v_2$ for Xe+Xe collisions at $\sqrt{s_{NN}}$=5.44 TeV. We have also tried to demonstrate the no. of constituent quark, $N_q$, and $m_T$ scaling of the elliptic flow. We find that $n_q$ scaling of $v_2$ is not observed for the identified hadrons. The $v_2$ results from Xe+Xe collisions have also been compared to Pb+Pb collisions at $\sqrt{s_{NN}}$ = 5.02 TeV. We find that flow of charged particles in (50-60)\% central collisions for Xenon nuclei is almost 30\% less than particle flow developed in  lead ion collisions, implying the important role the system size play in development of particle collective motion in relativistic heavy ion collisions.

\end{abstract}
 
\date{\today}
\maketitle 
\section{Introduction}
\label{intro}
The main goal of the ultra-relativistic heavy-ion collisions is to study the matter at high temperature and density where quantum chromodynamics predict the existence of the quark-gluon plasma (QGP)~\cite{QGP} under such extreme conditions. Relativistic Heavy Ion Collider (RHIC) and Large Hadron Collider (LHC) are the dedicated state-of-the-art experimental facilities to this end and are focused to understand the properties of QGP. Anisotropic flow as an observable of QGP, is caused by the initial asymmetries in the geometry of the system produced in any non-central collision, and plays an important role to understand the collective motion and bulk property of the QGP. The elliptic flow or azimuthal anisotropy ($v_{2}$), which is defined as the second-order Fourier component of the particle azimuthal distribution provides information about the equation of state, initial geometrical anisotropy and the transport properties of created QGP~\cite{v2}. The elliptic flow has been intensively  studied at RHIC and LHC for different systems like Au + Au, Cu + Cu and Pb + Pb at different center of mass energies from 7.7 GeV to 5.02 TeV. From many of the measurements from these previous experiments we have seen that $v_{2}$ has contributed significantly to the characterization of the system created in heavy-ion collisions as it is sensitive to the properties of the system at an early time of its evolution. Earlier RHIC experiments show that at low transverse momentum ($p_{T}$) region ($p_{T} <$ 2 GeV/c), $v_{2}$ follows the mass ordering i.e. higher mass particles having lower $v_{2}$ values~\cite{v2_star1, v2_star2, v2_star3, v2_star4, v2_phenix1, v2_phenix2, v2_phenix3, v2_phenix4}. Another important feature of $v_{2}$ is the the number of constituent quark (NCQ) scaling, where $v_{2}$ and $p_{T}$ of identified hadrons are divided by the number of constituent quarks ($n_{q}$). This scaling interprets the dominance of the quark degrees of freedom at early stages of the collision. Recently LHC shows similar mass ordering of $v_{2}$ at low-$p_{T}$ but it seems that $v_{2}$ does not follow the NCQ scaling at LHC energies~\cite{v2_ALICE1, v2_ALICE2} for intermediate or high momentum. It would be very interesting to study these properties of $v_{2}$ in QGP medium with varying spatial configurations and densities etc. of partons, which can be achieved by relativistic collisions of different species of ions with a large variation of mass number at same centre of mass energy. 

Recently LHC has collided $\rm Xe^{129}$ nuclei at $\sqrt{s_{NN}}$ = 5.44 TeV. Since the mass number of the $\rm Xe^{129}$ nuclei is roughly in middle proton and $\rm Pb^{208}$ nuclei, this can provide the unique opportunity to study the system-size dependence of elliptic flow at LHC energies. According to the recent hydrodynamical calculation~\cite{hydro_xexe}, $v_{2}$ is found larger by 25\% in Xe + Xe than in Pb + Pb collisions in 0-5\% centrality class but it is smaller by 10\% above 30\% centrality classes. An earlier prediction from A Multi-Phase Transport (AMPT) model suggests that the NCQ scaling will hold when we consider much smaller system than Pb + Pb, which has shown that the number of constituent quarks (NCQ), $n_q$ scaling holds for Si + Si collisions at $\sqrt{s_{NN}}$ = 2.76 TeV much better than that of Pb + Pb collisions at same energy~\cite{AMPT_Scaling}.  Hence, It is also expected that NCQ scaling will also hold true for $v_2$ in Xe+Xe collisions. As expected, this may also indicate the formation of partonic system in Xenon nuclei collisions similar to Au+Au or Pb+Pb collisions. Therefore the study of $v_{2}$ will be very interesting at LHC energies with smaller system size. In this article we have studied the $v_{2}$ of produced particles in Xe + Xe collisions at $\sqrt{s_{NN}}$ = 5.44 TeV using  A Multi-Phase Transport model (AMPT) with string melting version~\cite{AMPT1,AMPT3}. 

The paper is organized as follows. The next section briefly deals with the AMPT model we are using to calculate $v_{2}$ of charged particles. This is followed by section on results and their discussion. We then conclude our paper by summarizing our results and findings in the conclusions.

\begin{figure}[h]
\includegraphics[height=16em]{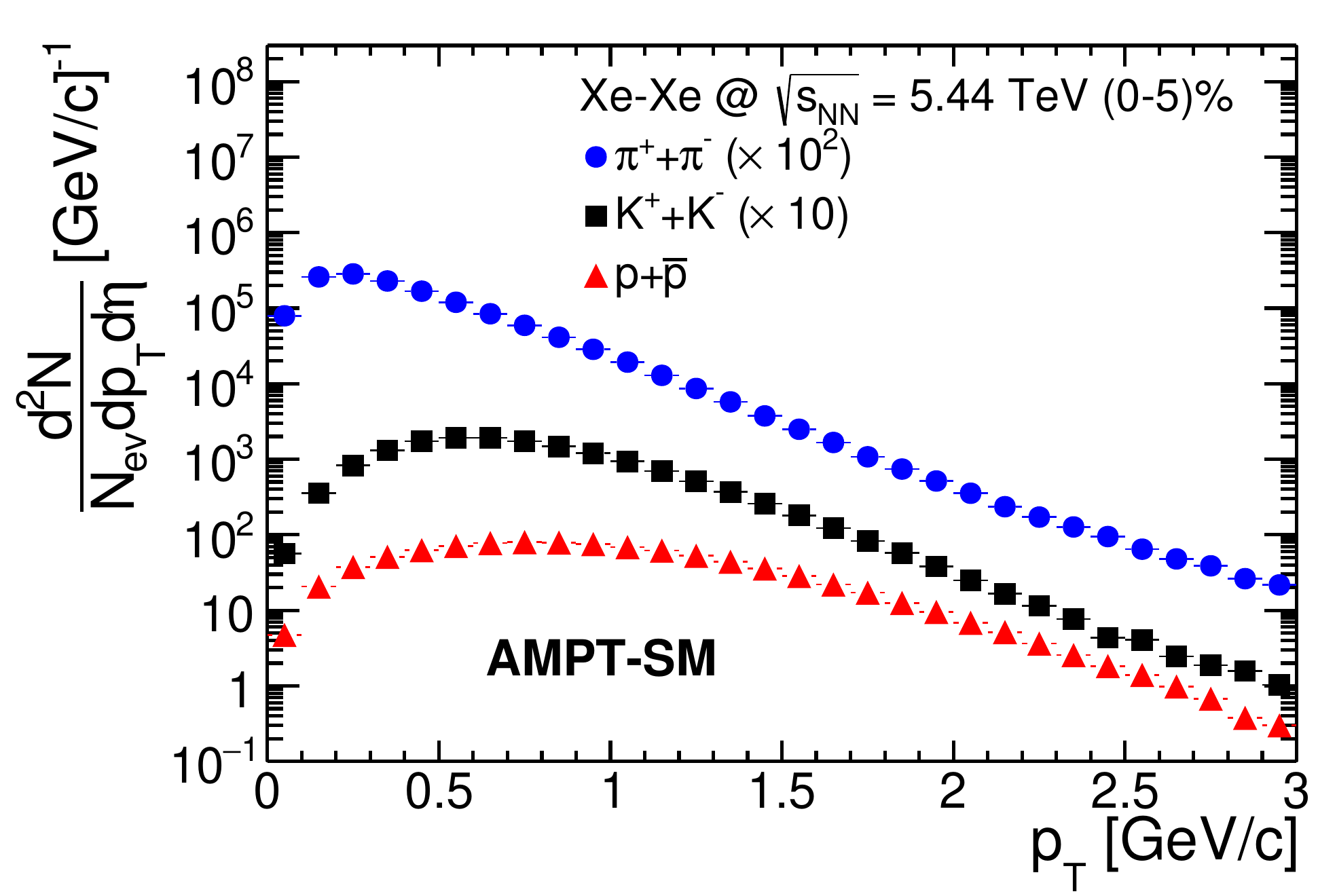}
\caption[]{(Color online) $p_T$ spectra of $\pi$, $K$ and $p$ for (0 - 5)\% centrality in Xe+Xe collisions. 
Circles are for pions, squares stand for kaons and triangles represent protons. The vertical lines in data points show the statistical uncertainties.}
\label{figptspectra}
\end{figure}

\begin{figure}[h]
\includegraphics[height=16em]{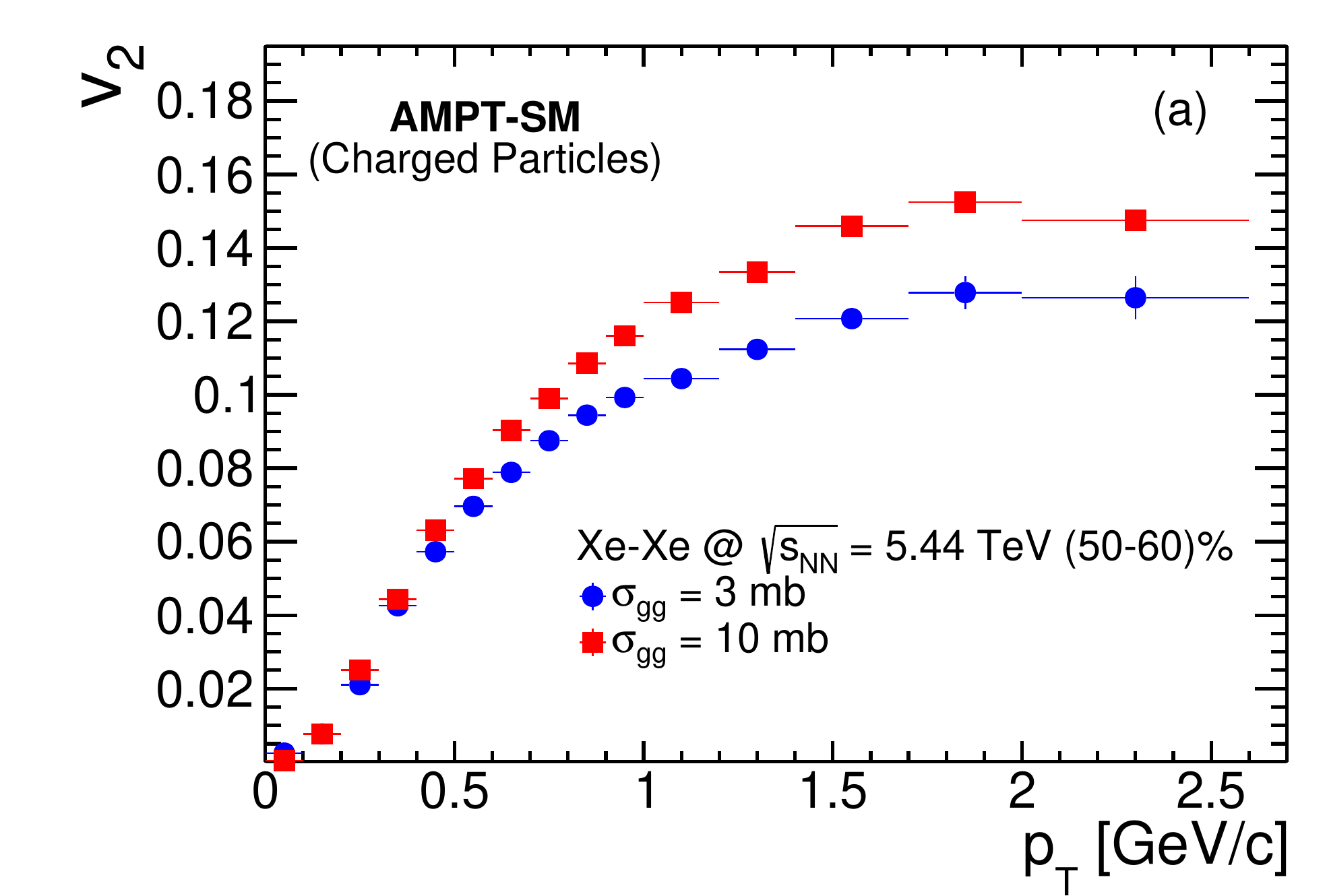}
\includegraphics[height=16em]{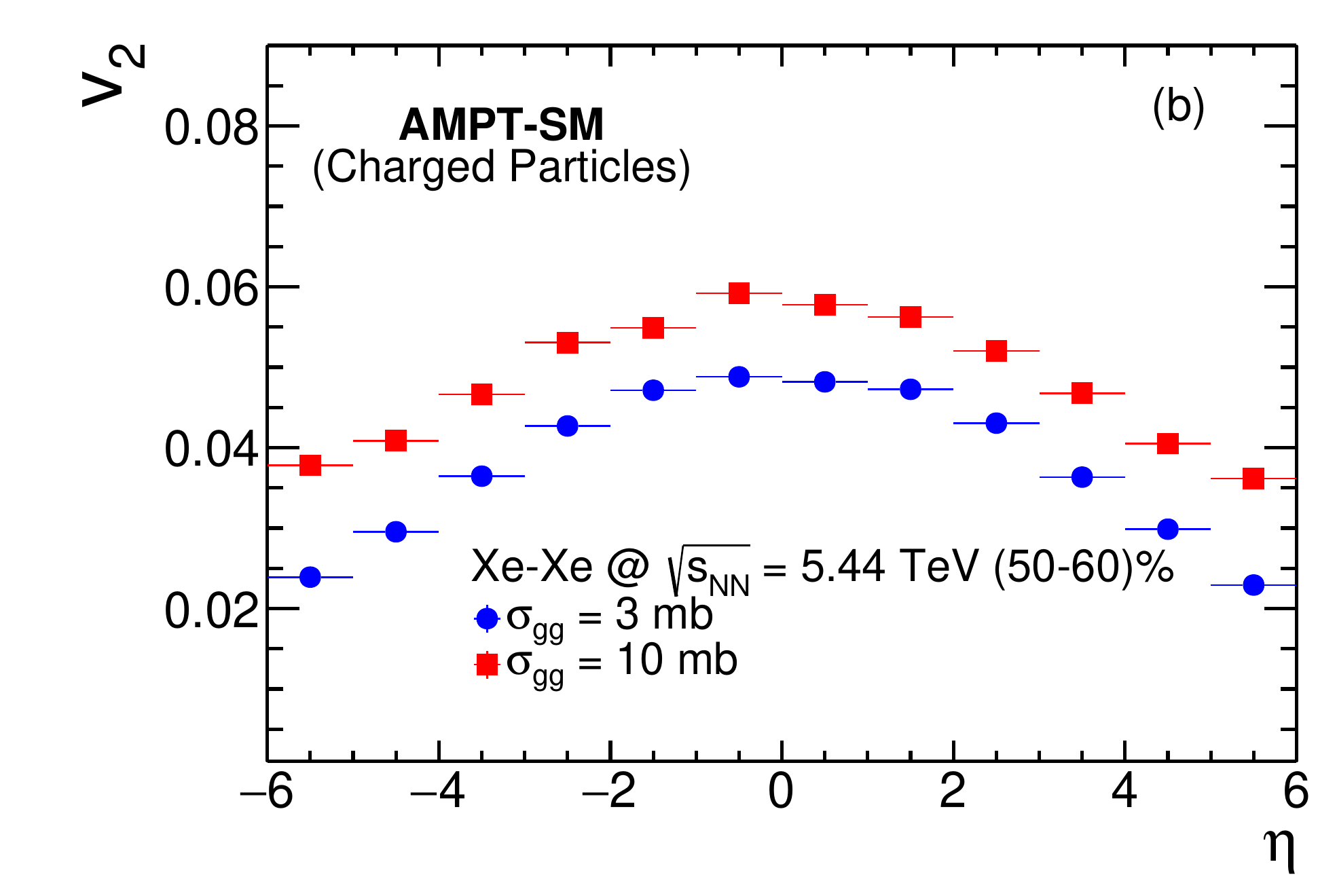}
\caption[]{(Color online) Charged particles $v_2$ vs. transverse momentum, $p_T$ (a), and pseudo-rapidity, $\eta$ (b). Different symbols are for $\sigma_{gg}$=3 mb and 10 mb.}
\label{fig3to10mb}
\end{figure}

\begin{figure*}[ht]
\centering
\includegraphics[scale=0.75]{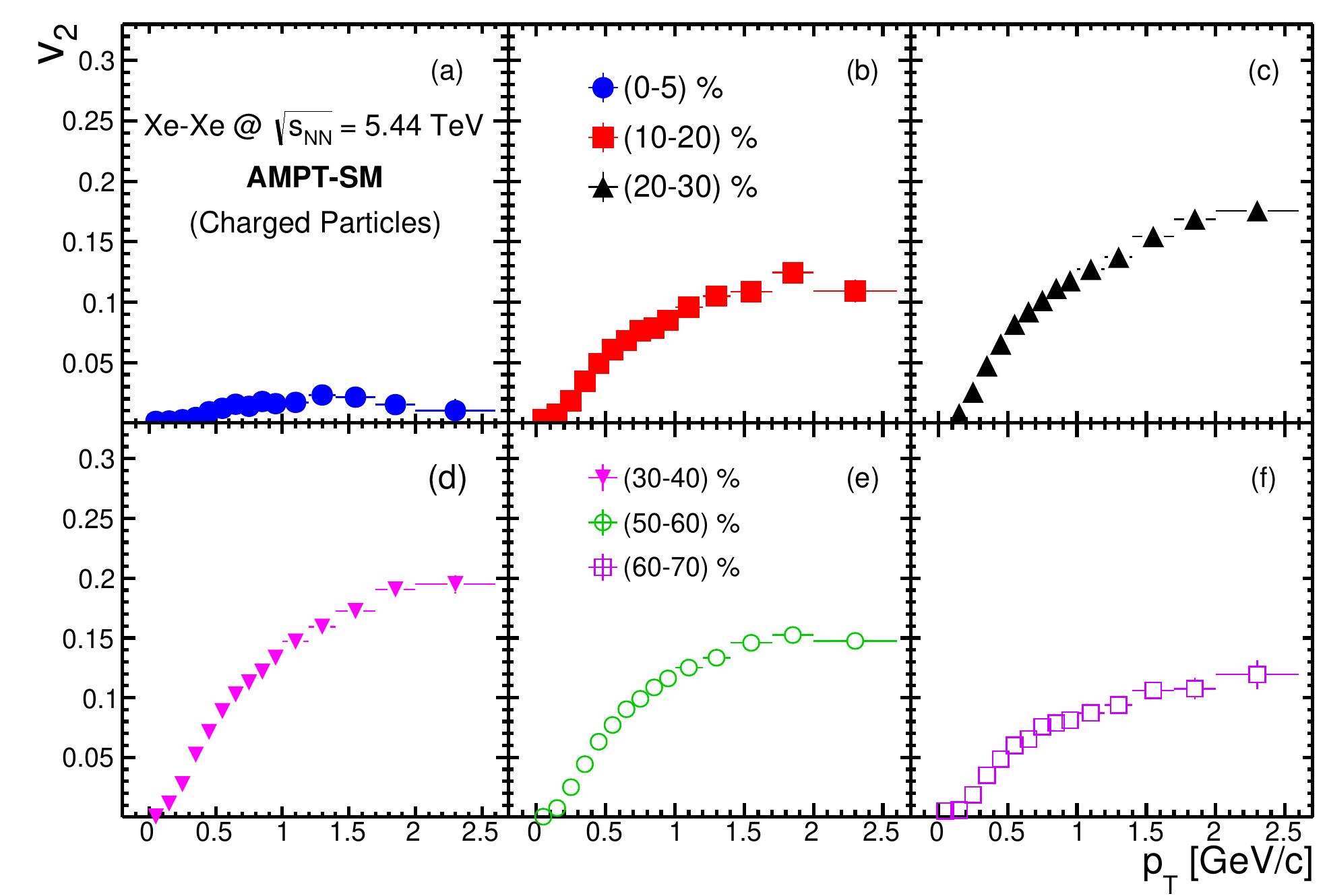}
\caption[]{(Color online) Charged particles $v_2$ as a function of transverse momentum, $p_T$ for different centrality classes. Different sections of the figure represent different centrality bins starting from (0-5)\% (a) to (60-70)\% (f).}
\label{figv2ptcentr}
\end{figure*}

\begin{figure}[h]
\includegraphics[height=16em]{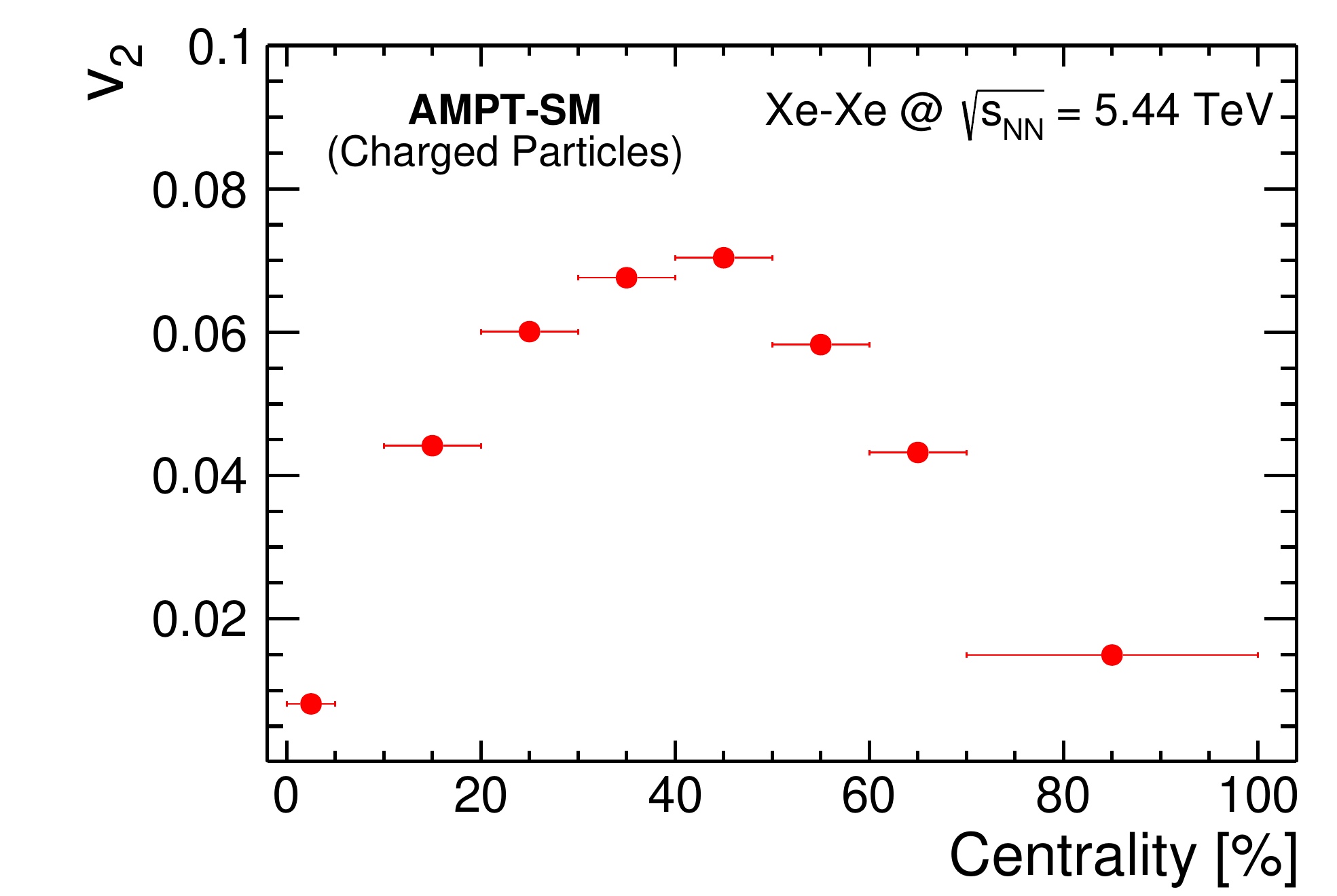}
\caption[]{(Color online) Charged particles $v_2$ vs. centrality for $|\eta| <$ 0.8. }
\label{figv2centr}
\end{figure}

\begin{figure}[h]
\includegraphics[height=16em]{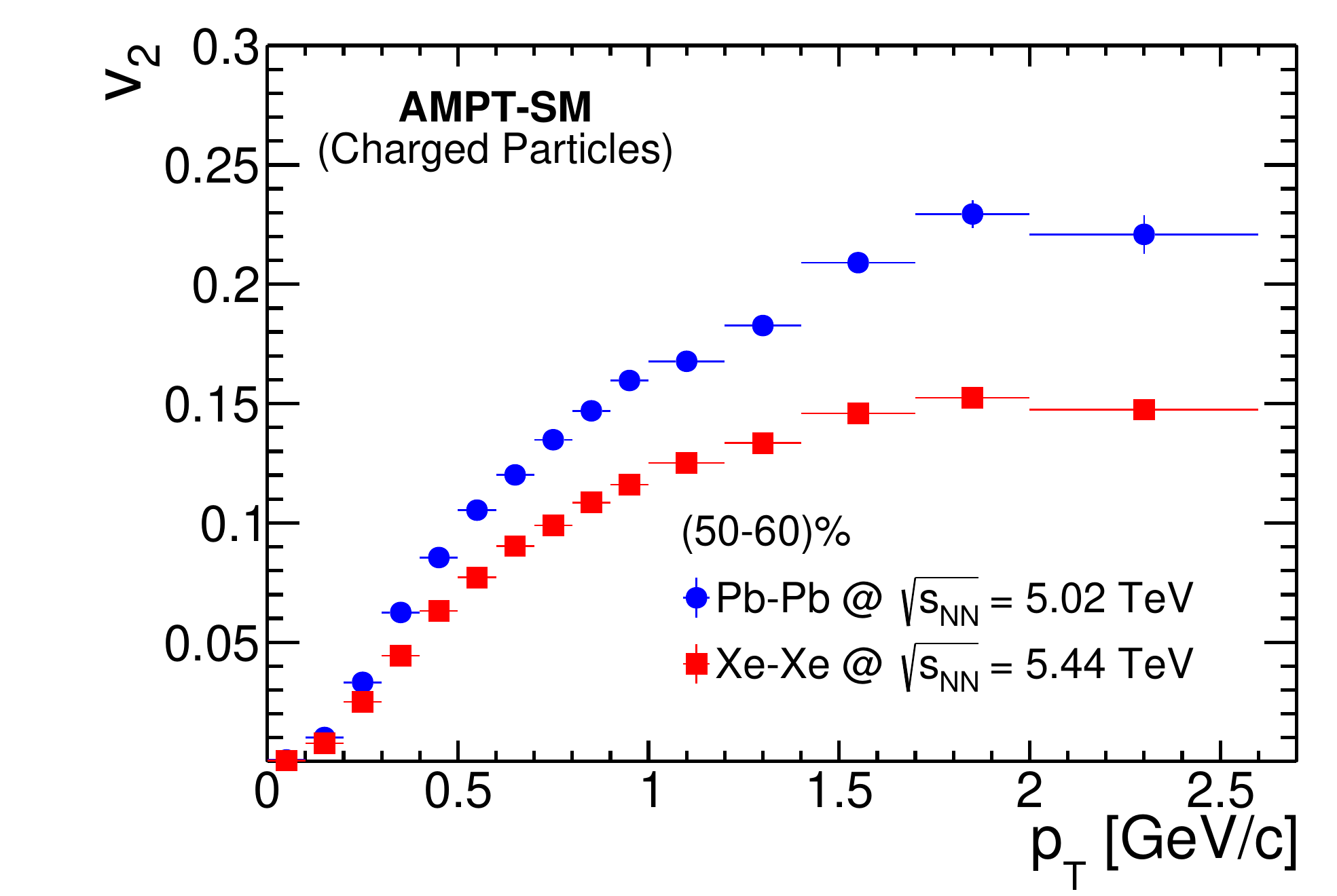}
\caption[]{(Color online) Charged particles $v_2$ vs. transverse momentum, $p_T$ for (50 - 60)\% centrality of Pb+Pb and Xe+Xe collisions. Circles are for Pb+Pb collisions at $\sqrt{s_{NN}}$ = 5.02 TeV and squares are for Xe+Xe collisions at $\sqrt{s_{NN}}$ = 5.44 TeV. The vertical lines in data points show the statistical uncertainties.}
\label{figXePb}
\end{figure}

\noindent
\section{A Multi-Phase Transport (AMPT) model}
\label{formalism}
AMPT is a hybrid transport model which contains four components namely, initialization of collisions, parton transport after initialization, hadronization mechanism and hadron transport~\cite{AMPT2}.  The initialization of the model follows HIJING model~\cite{ampthijing} and calculates the differential cross-section of the produced minijet particles in p+p collisions which is given by,

\begin{eqnarray}
\frac{d\sigma}{dp_T^2\,dy_1\,dy_2}=&K&\,\sum_{a,b}x_1f_a(x_1,p_{T1}^2)\,x_2f_2(x_2,p_{T2}^2)\nonumber\\
&\times&\frac{d\hat{\sigma}_{ab}}{d\hat{t}}\,,
\end{eqnarray}
where $\sigma$ is the produced particles cross-section and $\hat{t}$ is the momentum transfer during partonic interactions in p+p collisions. $x_i$'s are the momentum fraction of the mother protons which are carried by interacting partons and $f(x, p_T^2)$'s are the parton density functions(PDF). 
The produced partons calculated in $\rm p+p$ collisions is then converted into $\rm A+A$ and $\rm p+A$ collisions by incorporating parametrized shadowing function and nuclear overlap function using inbuilt Glauber model within HIJING. Similarly, initial low-momentum partons which are separated from high momenta partons by momentum cut-off, are produced from parametrized coloured string fragmentations mechanisms. The produced particles are initiated into parton transport part, ZPC~\cite{amptzpc}, which transport the quarks and gluons using Boltzmann transport equation which is given by,

\begin{eqnarray}
p^{\mu}\partial_{\mu}f(x,p,t)=C[f]
\end{eqnarray}

The leading order equation showing interactions among partons is approximately given by,

\begin{equation}
\frac{d\hat{\sigma}_{gg}}{d\hat{t}}\approx \frac{9\pi\alpha_s^2}{2(\hat{t}-\mu^2)^2}\,.
\end{equation}

Here $\sigma_{gg}$ is the gluon scattering cross-section, $\alpha_s$ is the strong coupling constant used in above equation, and $\mu^2$ is the cutoff used to avoid infrared divergences which can occur if the momentum transfer, $\hat{t}$, goes to zero during scattering. In the String Melting version of AMPT (AMPT-SM), melting of coloured strings into low momentum partons also take place at the start of the ZPC and are calculated using Lund FRITIOF model of HIJING. This melting phenomenon depends upon spin and flavour of the excited strings. The resulting partons undergo multiple scatterings which take place when any two partons are within distance of minimum approach which is given by $\displaystyle d\,\leq\,\sqrt{\sigma/\pi}$, where $\sigma$ is the scattering cross-section of the partons. In AMPT-SM, the transported partons are finally hadronized using coalescence mechanism~\cite{amptreco}, when two (or three) quarks sharing a close phase-space combine to form a meson (or a baryon). As of present, AMPT-SM uses three-momentum conservation and the invariant masses of the coalescing partons. 
The coalescence takes place using the following equation (for e.g. meson),
\begin{eqnarray}
\frac{d^3N}{d^3p_M}=g_M\int{d^3x_1d^3x_2d^3p_1d^3p_2\,f_q(\vec{x}_1,\vec{p}_1)}f_{\bar{q}}(\vec{x}_2,\vec{p}_2)\nonumber\\
                            \delta^3(\vec{p}_M-\vec{p}_1-\vec{p}_2)\,f_M(\vec{x}_1-\vec{x}_2,\vec{p}_1-\vec{p}_2).
\end{eqnarray}
Here $g_M$ is the meson degeneracy factor, $f_q$'s are the quark distributions after the evolution, and $f_M$ is the coalescing function commonly called Wigner functions~\cite{Greco:2003mm}.

The produced hadrons further undergo evolution in ART mechanism~\cite{amptart1, amptart2} via meson-meson, meson-baryon and baryon-baryon interactions, before final spectra can be observed. There is another default version of AMPT known as AMPT-Def where instead of coalescing the partons, fragmentation mechanism using Lund fragmentation parameters $a$ and $b$ are used for hadronizing the transported partons. However, it is believed that particle flow and spectra at the mid-$p_T$ regions are well explained by quark coalescence mechanism for hadronization~\cite{ampthadron1,ampthadron2,ampthadron3}. We have used AMPT-SM mode for our calculations. We have used the AMPT version 2.26t7 (released: 28/10/2016) in our current work.

Before we start discussing about $v_{2}$, it is important to have a look into the $p_{T}$ spectra of the produced particles such as $\displaystyle \pi, K, \text{and}\, p$. Fig.~\ref{figptspectra} shows $p_{T}$ spectra of $\displaystyle \pi, K, \text{and}\, p$ for (0-5)\% centrality. The error bars in the data points are the statistical uncertainties. The spectra of pions and kaons are multiplied by different constant factors to get a clear view of each spectrum. This provides a very good baseline to study the $p_{T}$ spectra of identified particles in experiment for Xe+Xe collisions.

It is worthwhile to mention that earlier studies of particle $v_2$ in Pb+Pb collisions with AMPT showed greater match with experimental data when large partonic scattering cross-section ($\sigma_{gg}$ $\approx$ 10 mb) is taken~\cite{amptsigma}. In fig.\ref{fig3to10mb}, we have shown azimuthal anisotropy or elliptic flow of charged particles, $v_2$, as functions of transverse momentum, $p_T$, and rapidity, $\eta$ for two values of scattering cross-sections, $\sigma_{gg}$ = 3 mb and 10 mb. As expected, results with 10 mb shows greater $v_2$ than 3 mb. In case of $p_T$ as variable, the flow increases more with transverse momentum in 10 mb scenario than 3 mb. While, taking rapidity, $\eta$, as the variable, the difference in 10 mb and 3 mb results can be seen as a constant multiplication factor, particularly in the central rapidity region. In the present work we have fixed $\sigma_{gg}$ = 10 mb as  cross-section for our calculations and calculated charged particle $v_2$. The Lund string fragmentation parameters $a$ and $b$ are kept fixed at their default values of 2.2 and 0.5/GeV$^2$, respectively. We will compare our results with the experimental data when it becomes available and further optimise the parameters.

The anisotropic flow can be characterized by the coefficients ($v_n$), which are obtained from a Fourier expansion of the momentum distribution of the charged particles and is given by,

\begin{eqnarray}
E\frac{d^3N}{d^3p}=\frac{d^2N}{2\pi p_Tdp_Tdy}\bigg(1+2\sum_{n=1}^\infty v_n \cos[n(\phi -\psi_n)]\bigg)\,,\nonumber\\
\end{eqnarray}

where $\phi$ is the azimuthal angle in the transverse momentum plane and $\psi_n$ is the n$^{\text{th}}$ harmonic event plane angle~\cite{v2eventplane}. In the current work elliptic flow is calculated with respect to the reaction plane by taking $\psi_{n}$ = 0, which implies event plane coincides with the reaction plane. Taking $n$ = 2 gives the second order harmonics in the expansion and its coefficient, $v_2$ is calculated to provide the measure of the elliptic flow or azimuthal anisotropy. For a given rapidity window the $v_{2}$ is defined as:

 \begin{eqnarray}
 v_{2} = \langle \cos(2\phi)\rangle
\end{eqnarray}

For non-central collisions the $v_2$ should be non-zero finite quantity.

\begin{figure}[h]
\includegraphics[height=16em]{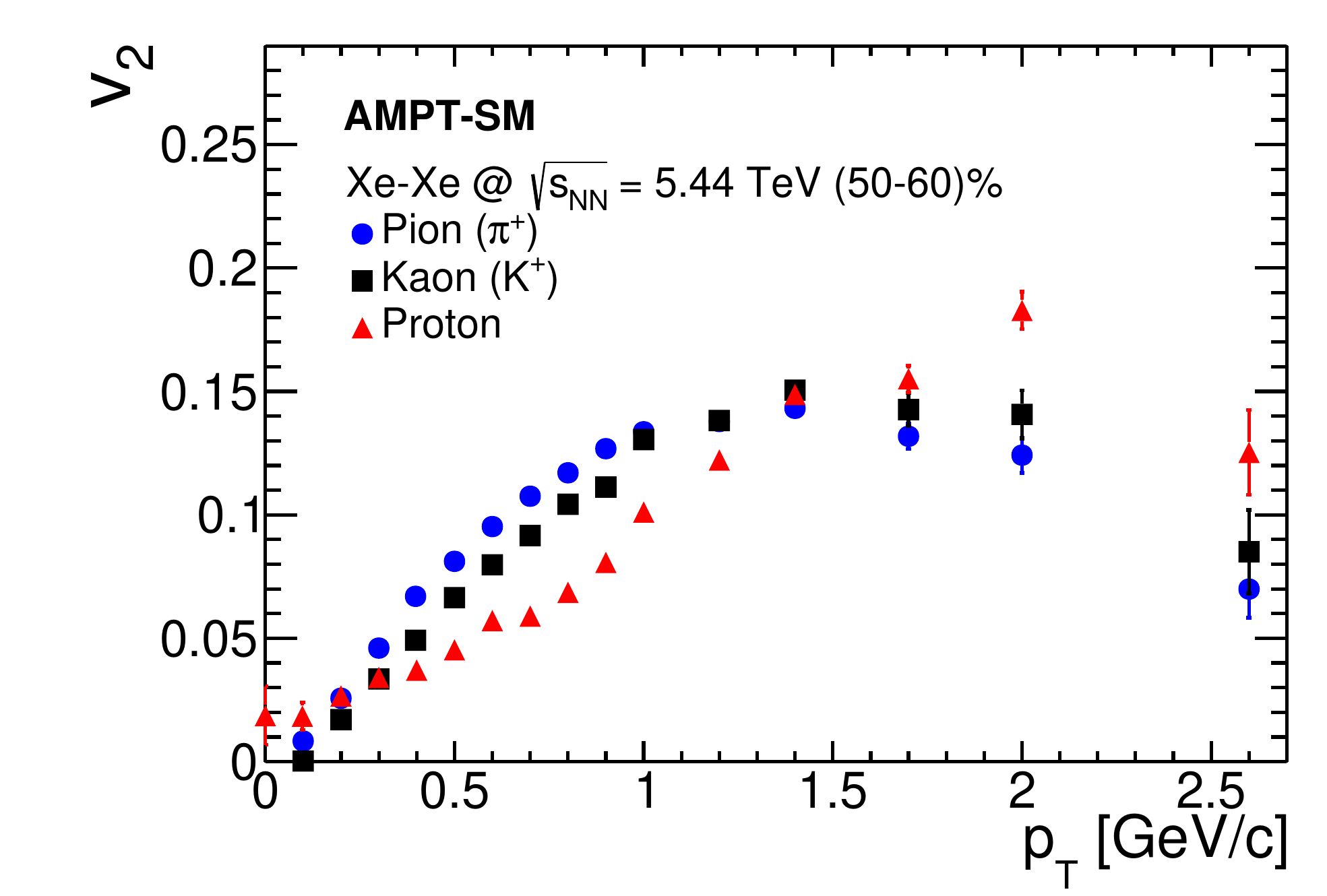}
\caption[]{(Color online) $v_2$ vs. transverse momentum, $p_T$  of $\pi$, $K$ and $p$ for (50 - 60)\% centrality in Xe+Xe collisions. 
Circles are for pions, squares stand for kaons and triangles represent protons. The vertical lines in data points show the statistical uncertainties.}
\label{figv2ptidp}
\end{figure}

\begin{figure}[ht]
\includegraphics[height=16em]{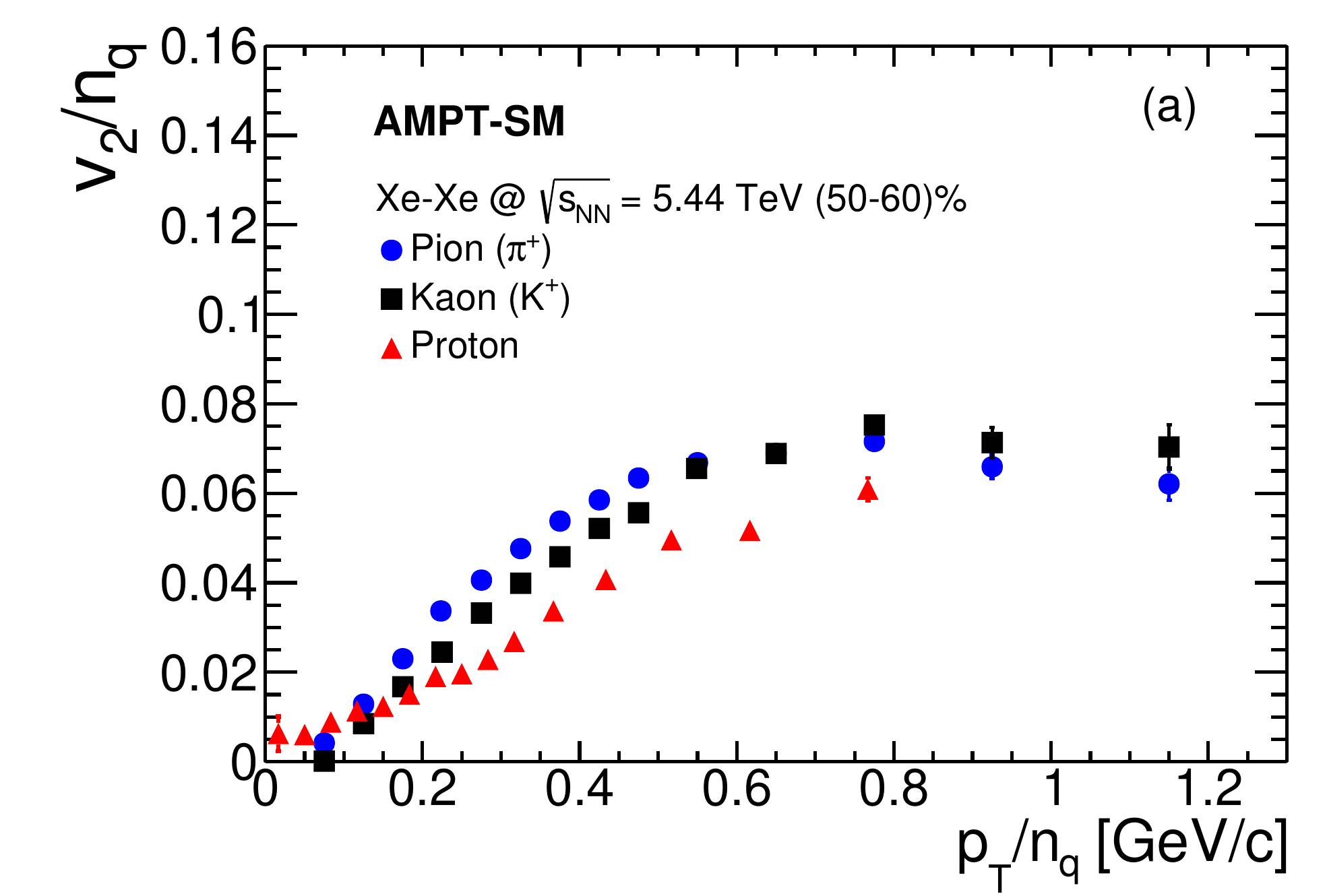}
\includegraphics[height=16em]{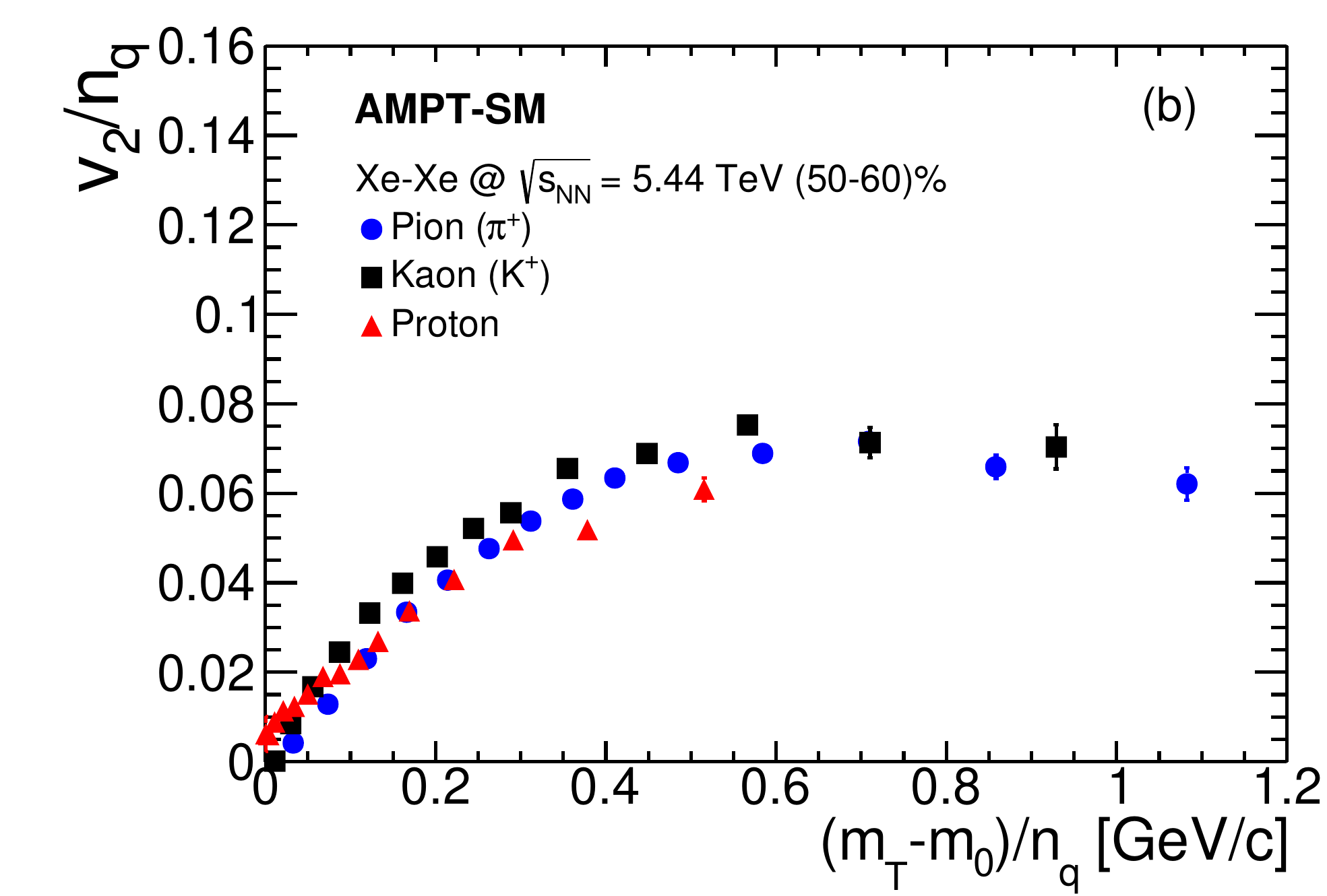}
\caption[]{(Color online) (a): $v_2/n_q$ as a function of $p_T/n_q$ for $\pi$, $K$ and $p$. (b):  $v_2/n_q$ as a function of $(m_T - m_0)/n_q$ for $\pi$, $K$ and $p$. Both plots are for 50-60\% centrality bin and different symbols represent different particles. The vertical lines in data points show the statistical uncertainties.}
\label{figv2ptnq}
\end{figure}

\section{Results and Discussions}
\label{results}
We have calculated charged particles elliptic flow or azimuthal anisotropy, $v_2$ for various centralities, namely, (0-5)\%, (10-20)\%, (20-30)\%, (50-60)\%, (60-70)\% centrality classes. Results are presented for 0 $< p_{T} <$ 2.7 GeV/c in midrapidity region ($|\eta| <$ 0.8). It is believed that Xe nucleus is moderately deformed. Earlier theoretical studies~\cite{hydro_xexe} on central collisions ($\approx$ (0-15)\%) of Xe nuclei have shown that incorporating deformation parameters cause about 15\% deviation in $v_{2}$ compared to non-deformed cases. Beyond 15\% centrality, it is claimed that the deformity has no discernible effects on particle spectra or other observables. In the present work as the first approximation, we haven't used any deformation for Xenon nuclei. Most of the results are shown for the above mentioned centralities for Xe+Xe collisions at $\sqrt{s_{NN}}$ = 5.44 TeV.

In fig.\ref{figv2ptcentr}, we have shown elliptic flow of charged particles for six different centralities. For the most central collision (0-5\%), the calculated $v_2$ is minimum. The elliptic flow increases as the centrality is increased as evident from the next three centralities [(10-20)\%, (20-30)\% and (30-40)\%] shown in the figure. However we see that the difference between (20-30)\% and (30-40)\% is small. In (50-60)\% onwards, we have a decreasing trend in the flow and smaller $v_2$, which continues to decrease for more peripheral collisions. We feel that this might be due to formation of smaller system in peripheral collisions at which although we have more geometrical anisotropy, the medium density is very small and probability of formation of collective motion decreases.

Fig.\ref{figv2centr} shows transverse momentum integrated $v_2$ with the centrality in the $p_T$ range mentioned above. As expected, a strong dependence of $v_{2}$ with centrality is observed. It is quite evident from this figure that $v_2$ increases from most central to mid-central collisions. Beyond that as we move towards peripheral collisions the flow decreases rapidly.  Similar behaviour of the charged particle elliptic flow has also been observed for Pb-Pb collisions at LHC energies~\cite{AMPT_PbPb}. Earlier calculations with hydrodynamical models give similar pictures of elliptic flow with centrality. However, our calculations underestimate the results of hydrodynamics roughly by (13-30)\%~\cite{hydro_xexe}. We will continue to optimise the AMPT parameters like $\sigma_{gg},\,a\, \text{and}\, b$ in accordance with the upcoming experimental data and study the centrality dependencies of particle flow for Xe nuclei collisions in our future works, which in turn will help in studying other observables of greater importance in order to characterize the systems formed in Xe+Xe collisions. 
However, the recent experimental data do not have results beyond 70\% centrality. The statistics are too low. 
With a purpose to study the effects of anisotropy at the peripheral collisions with sufficient statistics, we have used a big centrality range of 70-100\%. The $v_2$ shows a very small value at most peripheral collisions in AMPT scenario. We feel that although the spatial anisotropy is largest for the peripheral collisions, the medium density is too small to provide any collective flow effects and the interactions among partons are less.


In fig.\ref{figXePb}, we have compared $v_2$ of charged particles in Pb+Pb collisions at $\sqrt{s_{NN}}$ = 5.02 TeV with Xe+Xe collisions at $\sqrt{s_{NN}}$ = 5.44 TeV at midrapidity ($|\eta| <$ 0.8) using same configurations in the AMPT model. The available collisions energies for these two different systems are close to each other and hence we may be able to discern various properties of QGP which depend on system sizes. The dependence of elliptic flow on the system size is quite evident from the plot, as $v_2$ from Xe+Xe is always different from Pb+Pb collisions. For (50-60)\% centrality, the anisotropic flow is more in Pb system than that of Xe. The difference increases toward higher $p_{T}$ and around 30\% higher $v_2$ is observed in Pb system. We would like to re-iterate that inclusion of intrinsic deformations for Xenon nuclei in the calculations may change many of the above features of observables and their comparative studies. We will continue to investigate this particular feature and report in our future works.

Fig.\ref{figv2ptidp} shows $v_2$ for $\displaystyle \pi, K, \text{and}\, p$ upto $p_T$ = 2.7 GeV/c for (50-60)\% centrality. A clear mass dependency of hadrons' $v_{2} (p_{T})$ is observed for $p_{T} <$ 2 GeV/c as it has been observed in Pb-Pb collisions~\cite{v2_ALICE1, AMPT_PbPb} earlier. Lower mass particles have higher $v_{2}$. In particular, the pions and kaons show slightly more flow than proton for $p_T <$ 2.0 GeV/c, whereas afterwards proton takes over the pions and kaons. According to the hydrodynamical calculations there is an interplay between radial and elliptic flow which may play an important role in determining this mass-ordering of $v_{2}$ at low-$p_{T}$. For $p_{T} >$ 2 GeV/c $v_{2}$ is separated according to baryons and mesons. The quark-coalescence mechanisms,~\cite{ampthadron1,ampthadron2} which is able to explain flow at the intermediate or moderate $p_T$ ranges has been considered for hadronization in AMPT-SM model used in our calculations.

Within AMPT mechanisms, when a quark and an anti-quark are close in phase space with their momenta very close 
to each other, they coalesce to form a meson. Similarly when three quarks come closer in phase space, they recombine to form a baryon. 
Since we assume that coalescence mechanism should work in the intermediate $p_T$ region, the calculated 
elliptic flow of charged hadrons when divided by their coalesced constituent quark numbers, may exhibit 
$n_q$ scaling behaviour.
In fig.\ref{figv2ptnq} (a), we have shown number of constituent quarks, $n_q$, scaling of $v_2$ for $\pi, K, \text{and}\,p$ for (50-60)\% centrality. In present work, it is calculated as,
\begin{equation}
v_2^h(p_T)=n_q.v_2^q(p_T/n_q)\,,
\end{equation}
where $\displaystyle n_q$ is the number of constituent quarks for the charged hadrons considered in our work. 
However, the figure doesn't demonstrate the $n_q$ scaling behaviour. We observe that scaled $v_2$ for protons doesn't match with those of mesons, $\displaystyle \pi\,\text{and}\,K$. Such violation of $n_q$ scaling has been observed for charged particles $v_2$ in Pb+Pb collisions at the LHC energies~\cite{v2_ALICE1}. In fig.\ref{figv2ptnq} (b), we tried to demonstrate $n_q$ scaling of the particles flow, where instead of $p_T$ along x-axis, we have $\displaystyle (m_T-m_0)/n_q$, where $\displaystyle m_T=\sqrt{p_T^2+m_0^2}$. We find that the $v_{2}/n_{q}$ values as a function of $\displaystyle (m_T-m_0)/n_q$ show a scaling like behaviour which is similar to the observations for Pb+Pb collisions at LHC energies~\cite{v2eventplane}. However, we can not state conclusively that $n_q$ scaling is observed although something similar to $m_T$ scaling can be seen at low-momenta. The reason behind the failed $n_q$ scaling 
may be due to the partons forming baryons occupying different phase space as compared to partons forming mesons. 
The study of the correlation of the relaxation and freeze-out times with the flow may shed some light on the difference in the flow among quarks at the partonic level~\cite{bterta1}. This calls for a more deeper understanding of the relaxation of the bulk system of quarks and gluons.

\section{Summary}
\label{conclude}
Xe+Xe collision may provide us with a partonic system whose size is approximately in between those produced by $p+p$ and Pb+Pb collisions. The experimental results on anisotropic flow in Xe+Xe and Pb+Pb collisions should provide us with an opportunity to study system size dependence of $v_2$ at the approximately same collision energies. In the present work we have used AMPT-SM model to calculate charged particles' $v_2$ for Xe+Xe collisions at $\sqrt{s_{NN}}$=5.44 TeV at mid-rapidity region $|\eta| < 0.8$ in 0 $< p_{T} <$ 2.7 GeV/c. A strong centrality dependence of the $v_2$ is observed. $p_{T}$ differential $v_{2}$ is measured for identified particles such as, pions, kaons and protons and a clear mass ordering is observed for $p_{T} <$ 2 GeV/c. We have also tried to demonstrate the number of constituent quarks, $n_q$ of the elliptic flow as a function of $p_{T}/n_q$ as well as with respect to $\displaystyle (m_T-m_0)/n_q$. We find that $n_q$ scaling of $v_2$ is not observed for types of charged particles used for our studies. Also we haven't used any intrinsic deformations as a first approximation referring to earlier works, which suggest its small influence on observed particle spectra and $v_2$ beyond 15\% centrality collisions. However, most of the results are presented for (50-60)\% centrality bin where we do not expect the effects due to deformations. AMPT-SM underestimates the experimental results at higher $p_{T}$ ($p_{T} >$ 1 GeV/c) with default parameters. It suggests that one needs to tune the parameters to reproduce the experimental results. This study provides a baseline for the recent experimental results. This is discussed in the Appendix.  

\section{Acknowledgements}
The authors acknowledge the financial supports from ALICE Project No. SR/MF/PS-01/2014-IITI(G) of Department of Science and Technology, Government of India. ST acknowledges the financial support by DST-INSPIRE program of Government of India. This  research used  resources  of  the  LHC  grid  computing centers  at Variable Energy Cyclotron Center, Kolkata. 

\section{Appendix}
Recently ALICE has published the $v_{2}$ measurements of charged particles in Xe+Xe collisions at $\sqrt{s_{NN}}$ = 5.44 TeV~\cite{Acharya:2018ihu} using two-particle correlations based on cumulant method~\cite{Bilandzic:2010jr}. Fig.~\ref{figv2data} (in appendix) shows the comparison of $v_{2}$ vs $p_{T}$ between data and AMPT-SM model for 50-60\% centrality bin. It is found that AMPT result matches with data well at low $p_{T}$ region ($p_{T} <$ 1 GeV/c). Difference between data and AMPT increases as we go towards higher $p_{T}$. As there was no experimental results available before, we took the default parameter settings of AMPT-SM as it is mentioned in section II. It seems that to reproduce these latest experimental results, particularly at higher $p_{T}$ one needs to tune the parameters of AMPT-SM model. It should be noted here that the methods adopted in this paper and in the ALICE experimental paper for the measurement/estimation of $v_{2}$ are different. We have used the event plane method, whereas the experimental measurement uses the cumulant method. However, as discussed in Ref. \cite{Bilandzic:2010jr} both the methods give similar $v_{2}$-results.     

\begin{figure}[ht]
\includegraphics[height=16em]{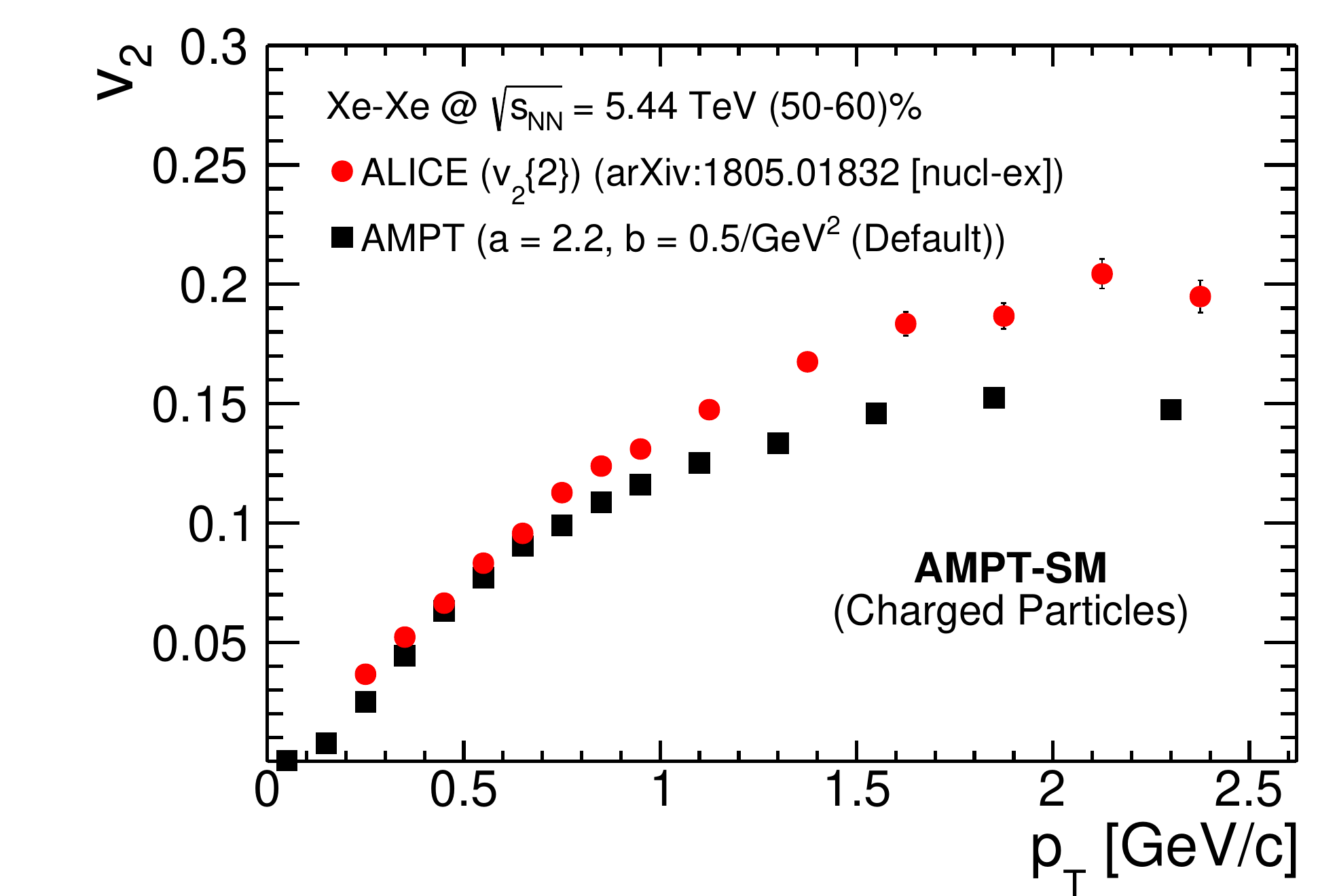}
\caption[]{(Color online) $v_2$ vs transverse momentum, $p_T$ of charged particles for ALICE data~\cite{Acharya:2018ihu} (circles) and AMPT-SM model (squares).}
\label{figv2data}
\end{figure}

\end{document}